# John von Neumann's 1950s Change to Philosopher of Computation

Steven Meyer
*smeyer@tdl.com*
August 25, 2020

**Abstract:** John von Neumann's transformation from a logician of quantum mechanics (QM) in the 1920s to a natural philosopher of computation in the 1950s is discussed. The paper argues for revision of the historical image of Neumann to portray his change to an anti formalist philosopher of computation. Neumann abandoned Hilbert's programme that knowledge could be expressed as logical predicates. The change is described by relating Neumann's criticism of Carnap's logicism and by discussing Neumann's rejection of the Turing Machine model of computation. Probably under the influence of the founders of modern physics in particular Wolfgang Pauli and Werner Heisenberg at the Advanced Study Institute, Neumann changed to a natural philosopher of computation. Neumann's writings from his development of the now almost universal von Neumann computer architecture are discussed to show his 1950s view of algorithms as physicalized entities. The paper concludes by quoting Neumann's statements criticizing mechanistic evolution and criticizing neural networks.

**Keywords:** John von Neumann, Von Neumann computer architecture, Philosophy of computation, Hilbert's programme, Turing Machine computation model, MRAM computation model, Rudolf Carnap, Anti formalism, Natural philosophy, Bohr's complementarity, P?=NP problem.

## 1. Introduction

John von Neumann's transformation from a logician of quantum mechanics (QM) in the 1920s to a natural philosopher of computation in the 1950s is best expressed by a story Neumann told about Wolfgang Pauli's criticism of formal mathematics. "If a mathematical proof is what matters in physics, you would be a great physicist." (Thirring[2001], p. 5). Neumann's transitioned from adherent of David Hilbert's programme that all knowledge could be axiomatized as predicate formulas in the 1920s to the first computer scientist is described. The von Neumann computer architecture (Aspray[1990]) anticipated current theoretical models of computation. Neumann also criticized formal automata and machine learning.

This paper argues that the wide spread popularity of computers and artificial intelligence has resulted in suppression of Neumann history. I argue that there is a need for much more detailed historical study of Neumann. This paper is a first step in the historical revision of the 1950s Neumann. In my view, to often the historical image of Neumann only includes his pre WW II work on formal quantum logic.

Neumann's transition to philosopher of computation became his dominant passion driving changes in his other philosophical positions. However, he remained an applied mathematician publishing earlier work even from the late 1940s and continued to work on problems from applied mathematics. One example is Neumann's lecture to the International Congress of Mathematics in 1954 titled "Unsolved problems in Mathematics" (Neumann[1954]). He considered infinitary problems in Hilbert spaces while mentioning that they did not correspond to physical reality (p. 241). Other contributions of the later Neumann such as his contribution to computer patents is not discussed because it requires legal expertise. Neumann's contribution to econometrics is not discussed.

## 2. Background

Kurt Godel's incompleteness results in the late 1920s (Godel[1931]) and discussions with physicists in the 1930s were some factors motivating Neumann's change. Neumann's transition is best expressed by his strong criticism of Carnap's predicate formula based conception of information (Kohler[2001]). The paper discusses Neumann's criticism of Carnap and argues that Kohler's explanation of the reasons for the rejection of Hilbert's programme is incorrect because in the 1950s while working with physicists at the Advanced Study Institute, Neumann rejected predicate formula based knowledge. The von Neumann computer architecture explicitly improves the Turing Machine model (TM).

## 3. Neumann Study of Natural Philosophy

Natural philosophy studies physical reality. Before the development of modern physics, there was no need for natural philosophy because Newtonian physics described a fixed and causal world in terms of Newton's laws. Max Planck and Albert Einstein called themselves natural philosophers because their study of physics involved studying philosophical concepts and scientific methods. It was clear in the late 19th century that Newtonian physics was inadequate in some areas. The new modern physics redefined philosophical concepts such as causality and simultaneity and modified empirical studies to include methodological study of experiments (Heisenberg[1958] Chapter 6 for a more detailed description of this change).

The philosophical changes in physics remain the subject of debate. One important debate related to computation involves search for one unified physical theory that was probably due to Einstein. Heisenberg provided a modern explanation of unified physical theories unified by mathematical groups (Heisenberg[1958], pp. 105-107). In contrast, David Bohm held the view view that nature is an infinity of different qualitative realities (Bohm[1957]).

John von Neumann undertook the study of natural philosophy as part of his development of the modern digital computer starting in the early 1940s. Neumann was influenced by the founders of modern physics and physicists were influenced by Neumann in their concepts and methods used in calculating field properties. Neumann's natural philosophy of computation is usually studied in the computer science (CS) area as operations research. In contrast to Neumann's view, the other CS research programme is based on mathematical logic often expressed as the Church Turing Thesis. Logic based CS is now predominant.

## 4. Neumann's Criticism of Carnap's Conception of Information Shows Change

Eckehart Kohler in his excellent and detailed paper "Why von Neumann rejected Carnap's Dualism of Information Concepts" (Kohler[2001]) argues that Neumann's (and Pauli's) very harsh criticism of Carnap was incorrect because Neumann wrongly assumed information only has physical meaning. Both Neumann and Pauli were so sure of their criticism that they recommended Carnap not publish his study of information. Neumann's criticism shows his changed views concerning logic and empiricism.

Neumann's opposition to Carnap's argument that information has a formal logical component as well as an experimental physical component provides the clearest explication of his changed philosophy. Kohler's mistake is that starting after world II physicists and applied mathematicians including Neumann rejected the very idea that formal sentence based logic can describe reality. Their criticism was not that Godel's results (Godel[1931]) made formula based propositional logic no longer absolute, but that logic failed as a method for describing the world.

By the early 1950s Neumann viewed the world as empirical for which growth of knowledge required experiments. I interpret Kohler and Carnap as arguing that Hilbert's programme could still work in conjunction with the dual idea that information has both a logical element and an empirical element (p. 117). Both Carnap and Kohler are attempting to save Hilbert's programme that they problem shift to be interpreted semantically. Kohler characterized Carnap's logical component of information as 'logical and mathematical sentences.' (p. 98) This is why Neumann attacked Carnap's dualism.

Neumann's rejection of his 1920s belief in Hilbert's programme can be seen from the advice Neumann gave Claude Shannon to use the term entropy for one of the functions involved in Shannon's definition of information in spite of Shannon's misgivings (p. 105). The advice was given around 1949 before Neumann's anti formalist views were fully developed, but shows Neumann's shift to empiricism. Shannon avoided the issue of the meaning of information by explaining it as simply a type of mathematical coding function.

The background change to anti-formalism did not just occur at the Institute of Advanced Study in Princeton, but was common among physicists and applied mathematicians after WWII. Formal propositional logic was viewed as problematic not only because systems of formal sentences could be used to mechanically derive inconsistent results (Godel1931], see also Paul Finsler's earlier results that were not tied to Russell's propositional calculus. Finsler[1996], Breger[1992]), but also because it was believed that knowledge not derivable from formal systems (from axioms plus formal propositions) exists. One obvious example that can not be understood in terms of formal mathematics is instantaneous wave function collapse in quantum mechanics.

After Neumann's death in the 1960s, empirical alternatives to logicism were developed. Finsler proved that the continuum hypothesis is true by defining a continuum that was axiomatically defined, but different from the standard continuum definition (Finsler[1968]). Einstein expressed earlier the viewpoint of the founders of modern physics in his 1921 lecture on geometry and experience (Einstein[1921]). Einstein argued that formal mathematics (propositional logic based rationality) was incomplete in a physical sense. However, Einstein expressed a contrary view in other writing. George Polya who was Neumann's teacher encouraged Imre Lakatos to solidify the ideas expressed by Neumann and Wolfgang Pauli that possibly actually originated with Polya. The idea in the 1960s was called quasi-empirical mathematics. Polya encouraged Lakatos to rewrite his thesis *Proofs and Refutations.* as a simpler book because it would solidify empirical mathematics in place of logicism (Polya[1975]).

## 5. Neumann Conception of Computation Closest to MRAM Model

Starting in the early 1940s, Neumann became focused on the possibilities of digital computers (Aspray[1990] for a detailed history). There is currently no evidence of Neumann's explicit discussion of his philosophy of computation. However, it is known that Neumann discussed his thinking about abstract properties a computer architecture should have and discussed problems with neural networks as models for computation. Neumann rejected that TM model because it used logics rather than physical properties (Kohler[2001], p. 104). Kohler explains Neumann's view of logic (algorithms) as physicalized entities (p. 116). Neumann's explicit rejection of formal neural networks also sheds light on his computational philosophy (Aspray[1990], note 94, p. 321).

In the early 1950s, the possibility was considered that computational errors in constructed computers were physically inherent conceptually similar to entropy. Since Neumann worked with Wolfgang Pauli and Werner Heisenberg who developed the Bohr interpretation of QM, the early view of errors may have expressed their view of Bohr's complementarity. It turned out the errors were circuit design mistakes and environment background caused errors that were repairable using error correcting codes.

However, the problem of inherent computational errors has recently re-emerged as errors in quantum computers. The errors (complementarity between classical macro physics and quantum micro physics) may actually represent new physical reality that can be measured with the new methods of cold atom physics (Monroe[2018]). It is possible that Neumann working with Pauli and Heisenberg anticipated quantum computing. I think Arthur Fine's characterization of complementarity as only a conceptual device is incorrect (Fine[1996], pp. 20, 21, 124).

This paper provides an analysis of Neumann's thinking by discussing a formalization of the Neumann computer architecture using a modern abstract model called MRAMs (random access machines with unbounded cell size and unit time multiply) first studied by Hartmanis and Simon (Hartmanis[1974][1974a]). MRAM's are the closest studied model to the Neumann computer architecture. Neumann explicitly listed the MRAM model properties in his list of requirements for his now almost universally used von Neumann computer architecture.

In contrast to TMs that have an unbounded number of bounded size unary representing memory cells, Neumann assumed that a computer would have a finite number of binary encoded unbounded size memory cells (computers need to be built large enough for the given problem). Neumann also argued that some sort of intuition needed to be built into programs instead of brute force searching (Aspray[1990], p. 62). Unlike TMs, Neumann's computer design provides bit selects and indexing.

## 6. Consequences of Lack of Study of Neumann's 1950s Change

There are modern consequences of Neumann's philosophy and indirectly lack of study of the 1950s Neumann. In the MRAM model, the P?=NP problem does not exist (or the answer is that there is no difference between the class of problems solvable by non deterministic guessing versus the class solvable by deterministic searching) (Meyer[2016]). Many modern philosophical questions assume implicitly that non deterministic TMs are more powerful than deterministic machines. For example, Shor's quantum computation algorithms assume P!=NP (Shor[1996]).

## 7. Examples of Neumann's 1950s Thinking

An interesting story related to Neumann's philosophy is that advocates of the importance of the P?=NP problem (does guessing speed up computation by more than a polynomial bound) found a letter in the Godel Archive to Neumann that they interpret as Godel supporting the importance of the P?=NP problem. However, in a letter (unfortunately undated) from Neumann to Oswald Veblen relating to an Institute of Advanced Study permanent appointment for Godel, Neumann shows skepticism toward Godel's later work (Hartmanis[1989] for the Godel P?=NP letter. Neumann[2005], p. 276 for the Neumann letter).

To me the most important consequence of Neumann's transformation is that he correctly understood that TMs are very weak (slow) computing machines. It is true that TMs are universal in the Church Turing sense (Copeland[2015]). Any TM or MRAM can calculate anything recursively computable in the Church Turing sense (for background the question 'are two regular expressions equivalent' is outside NP but computable). The thesis is often expressed as any TM can simulate any computing device. The problem is that Neumann understood the importance of computational efficiency so TM's lack of computational power is a negative factor. TMs are slower than von Neumann computers because the von Neumann architecture's indexed data structures obviate the need for non deterministic TM guessing.

Neumann also criticized other models of computation such as primitive automata and neural networks. The criticism is based on rejection of the relevance of predicate logic formulas in describing reality. It is not clear if Neumann rejected the Church Turing thesis. He seems not to have discussed it. The paper explains this skepticism by analyzing Neumann's dislike of Carnap's philosophy. Here are two specific examples of Neumann's criticism of automata.

## 7.1 Criticism of mechanistic evolution

**He (Neumann) led the biologist to the window of his study and said: 'Can you see the beautiful white villa over there on the hill? It arose by pure chance. It took millions of years for the hill to be formed; trees grew, decayed and grew again, then the wind covered the top of the hill with sand, stones were probably deposited on it by a volcanic process, and accident decreed that they should come to lie on top of one another. And so it went on. I know, of course, that accidental processes through the eons generally produce quite different results. But on this one occasion they led to the appearance of this country house, and people moved in and live there at this very moment.' (story told in Heisenberg[1971] p. 111)**

## 7.2 Low complication levels qualitatively different than high levels

**The insight that a formal neuron network can do anything which you can describe in words is a very important insight and simplifies matters enormously at low complication levels. It is by no means certain that it is a simplification on high complication levels. It is perfectly possible that on high complication levels the value of the theorem is in the reverse direction, namely, that you can express logics in terms of these efforts and the converse may not be true. (Neumann quoted in Aspray[1990], note 94, p. 321)**